\documentclass[prl,aps,a4paper,10pt,twocolumn,superscriptaddress,amsmath,amssymb,showpacs]{revtex4}
\usepackage{bm}
\usepackage{epsfig}
\usepackage{amsmath}

\renewcommand{\v}[1]{{\bf #1}}
\newcommand{\bpm}{\begin{pmatrix}}
\newcommand{\epm}{\end{pmatrix}}
\newcommand{\ba}{\begin{eqnarray}}
\newcommand{\ea}{\end{eqnarray}}
\newcommand{\nn}{\nonumber \\}

\begin{document}

\title{Fate of Topology in Spin-1 Spinor Bose-Einstein Condensate}

\author{Yun-Tak Oh}
\affiliation{Department of Physics, Sungkyunkwan University, Suwon
440-746, Korea}
\author{Panjin Kim}
\affiliation{Department of Physics, Sungkyunkwan University, Suwon 440-746, Korea}
\author{Jin-Hong Park}
\affiliation{Department of Physics, Sungkyunkwan University, Suwon 440-746, Korea}
\author{Jung Hoon Han}
\email[Electronic address:$~~$]{hanjh@skku.edu}
\affiliation{Department of Physics, Sungkyunkwan University, Suwon
440-746, Korea} \affiliation{Asia Pacific Center for Theoretical
Physics, POSTECH, Pohang, Gyeongbuk 790-784, Korea}
\begin{abstract}
One of the excitements generated by the cold atom systems is the possibility to realize,
and explore, varied topological phases stemming from multi-component nature of the
condensate. Popular examples are the antiferromagnetic (AFM) and the
ferromagnetic (FM) phases in the three-component atomic condensate with effective spin-1,
to which different topological manifolds can be assigned. It follows, from consideration
of homotopy, that different sorts of topological defects will be stable in each manifold.
For instance, Skyrmionic texture is believed to be a stable topological object in
two-dimensional AFM spin-1 condensate. Countering such common perceptions,
here we show on the basis of a new wave function decomposition scheme that there is no
physical parameter regime wherein the temporal dynamics of spin-1 condensate can be
described solely within AFM or FM manifold. Initial state of definite topological number
prepared entirely within one particular phase must immediately evolve into a mixed state.
Accordingly, the very notion of topology and topological stability within the
sub-manifold of AFM or FM become invalid. Numerical simulation reveals the linear Zeeman
effect to be an efficient catalyst to extract the alternate component from an initial
topological object prepared entirely within one particular sub-manifold, serving as a
potential new tool for ``topology engineering" in multi-component Bose-Einstein
condensates.
\end{abstract}
\date{\today}
\maketitle

{\it Introduction.}- Much of the novelty and surprise of the Bose-Einstein condensate
(BEC) of cold atoms arises from its multi-component nature~\cite{ueda-review}. Different
hyperfine states of the atom form components of ``spin", described by the wave function
$\Psi_F$ with $(2F+1)$ complex components in the case of spin-$F$ spinor condensate.
Being a dilute liquid state of matter, the dynamics of spinor BEC is subject to the
hydrodynamic description, in an extension of the earlier, popular approach to describe
scalar superfluid $^4$He~\cite{khalat}. The $2\times(2F+1)$ degrees of freedom in
spin-$F$ condensate may be partitioned into the density $\rho$, the overall phase
$e^{i\theta}$, and the remainder that may collectively be dubbed the internal degrees of
freedom. In the two-component $F=1/2$ case such counting argument leaves 4-2=2 internal
degrees of freedom corresponding precisely to the two orientation angles of the spin unit
vector $\v d =(\Psi^\dag_{F=1/2} \bm \sigma \Psi_{F=1/2}) /\rho$, $\bm \sigma =
(\sigma_x, \sigma_y, \sigma_z )$ being the Pauli matrices. The governing dynamics of
$F=1/2$ cold atoms is that of orientation fluctuation of the unit spin vector $\v d$, in
addition to the characteristic density and phase fluctuations of the
superfluid~\cite{lamacraft,refael}.

For spin-1 condensate the $6-2=4$ internal degrees of freedom are
harder to characterize and several attempts ranging from Schwinger
boson (SB) construction~\cite{refael} to (spin+nematic)
decomposition~\cite{yukawa-ueda} exist in the literature. These
approaches are complementary ways to faithfully characterize the
overall manifold of spin-1, three-component wave functions.
Meanwhile, Ho~\cite{ho} early on identified the two extreme limits
of the spin-1 manifold dubbed antiferromagnetic (AFM, also called
``polar") and ferromagnetic (FM) phases, obtained as the Euler
rotation $\mathcal{U}(\alpha, \beta, \gamma)= e^{-i F_z \alpha}
e^{-i F_y \beta} e^{-i F_z \gamma}$ of the spinors $\bpm 0 & 1 & 0
\epm^T$ and $\bpm 1 & 0 & 0 \epm^T $, respectively. The three
matrices $(F_x , F_y , F_z)$ form components of the spin-1 vector
$\v F$ in the angular momentum basis: $F_z = \mathrm{diag}(1,0,-1)$.
Discussions of topology in spin-1 condensate are typically built
around a specific sub-manifold~\cite{ho,zhou,ueda-review}. For
instance in the AFM phase the corresponding manifold ${\cal
M}_\mathrm{AFM} = ( U(1)\times S^2 )/\mathbb{Z}_2 $ was predicted to
support stable topological configurations of integer winding number
by the second homotopy rule $\pi_2 ({\cal M}_\mathrm{AFM} )
=\mathbb{Z}$; these are the two-dimensional Skyrmions.

Recently, Shin's group~\cite{shin} demonstrated the creation of a
single Skyrmion configuration in the antiferromagnetic spin-1
$^{23}$Na condensate. Rather surprisingly, the Skyrmion state they
created appeared to evolve smoothly into the non-textured state
without the topological object ever tunneling out of the condensate.
The standard wisdom, on the other hand, would be that a soliton of
definite topological charge is able to decay only through tunneling,
or annihilation with another soliton of opposite topological charge.
Numerical simulation~\cite{xu-han} based on time-dependent
Gross-Pitaevskii (GP) equation showed that an initial state of fully
antiferromagnetic Skyrmion texture evolves continuously into a mixed
phase of both AFM and FM components, under all reasonable ranges of
interaction strengths. Once there occurs substantial mixing, the
manifold on which the homotopy consideration rests becomes invalid,
let alone the topological number defined on that manifold.  One may
compare it to the case in two dimensions where the spins are
arranged as a vortex structure. Inhabitants of the two-dimensional
world, versed in homotopy theory yet mistakenly assuming the spins
to be completely of planar character, might anticipate the
robustness of the vortex object based on known homotopy result
$\pi_1 (S^1 ) = \mathbb{Z}$, since the planar spins have the
manifold of the unit circle $S^1$. Only, the spins were in fact of
Heisenberg character, or an element of the two-sphere $S^2$, and
could be smoothly deformed into a uniform state through rotation
{\it  out of the plane}. The observers are no longer shocked at the
smooth disappearance of the vortex knowing that the homotopy group
$\pi_1 (S^2 )$ is trivial.

The case of topological texture in the AFM manifold, as we demonstrate below, is analogous. It can
disappear gradually by rotating partly into a larger manifold involving the FM component. In fact,
we prove a stronger statement: the temporal dynamics defined exclusively within the AFM or the FM
manifold simply does not make sense. \textit{A given initial state prepared at time $t=0$ in a
particular sub-manifold inevitably has to evolve into a mixed state regardless of the strengths of
interactions.} A similar loosening of the original topological manifold occurs in the core region
of a vortex~\cite{ueda-vortex}. A vortex characterized by the non-trivial homotopy $\pi_1 ({\cal
M}_\mathrm{AFM} ) = \mathbb{Z}$ necessitates the presence of a singular core, with a diverging
kinetic energy that eventually overcome the spin-spin interaction. With the reversal of energy
hierarchy, the relevant order parameter space is no longer ${\cal M}_\mathrm{AFM}$ but a larger
manifold of $S^5$, whose trivial homotopy $\pi_1 (S^5 ) =0$ displaces the very notion of
topologically stable vortex-like object in the core region. Thus the notion of a vortex for spin-1
BEC is only well-defined in the far region where the kinetic energy is weak.

In our own discussion the breakdown of topology occurs in the temporal direction, and not due to
the energetic reason as discussed above for the vortex core. Rather, the root of the breakdown lies
with the impossibility to define sensible temporal dynamics within a sub-manifold ${\cal
M}_\mathrm{AFM}$ or ${\cal M}_\mathrm{FM}$ in the first place. The statement applies regardless of
the interaction strengths. To our knowledge, such breakdown in the notion of topology in the
temporal direction has never been suspected before.
\\

{\it New formulation of spin-1 condensate dynamics}.- General spin-1 condensate wave
function can be decomposed as $\Psi = \sqrt{\rho}e^{i\theta}\bm \eta$, where $\bm \eta$
is the unit-modulus complex CP$^2$ field satisfying $\bm \eta^\dag \bm \eta = 1$.
Particular subsets of CP$^2$ fields obtained by Euler rotations of the basis spinors
$\bpm 0 & 1 & 0 \epm^T$ and $\bpm 1 & 0 & 0 \epm^T $ are called AFM (subscript A) and FM
(subscript F):

\ba \bm \eta_A\!=\! \bpm -{1\over\sqrt{2}} e^{-i\alpha} \sin \beta
\\ \cos\beta \\ {1\over\sqrt{2}} e^{i\alpha} \sin \beta \epm, ~
\bm \eta_F \!=\! e^{-i\gamma}
\bpm e^{-i\alpha} \cos^2 {\beta \over 2 } \\
{1\over\sqrt{2}} \sin \beta \\ e^{i\alpha} \sin^2 {\beta \over 2 }
\epm . \label{eq:eta1-eta2} \ea
They represent two extreme situations of full, $(\bm \eta^\dag_F \v
F \bm \eta_F )^2 = 1$, and zero, $\bm \eta^\dag_A \v F \bm \eta_A =
0$, magnetizations, respectively. A spinor state of partial
magnetization can be constructed generally as a linear combination
of $\bm \eta_A$ and $\bm \eta_F$, with a mixing angle $\delta$:

\ba \Psi_{F=1} = \sqrt{\rho} e^{i\theta} \left( z_A  \bm \eta_A  + z_F \bm \eta_F \right)
\label{eq:our-Psi}\ea
with a mixing angle $\delta$: $(z_A , z_F ) = (\cos \delta/2, \sin \delta /2)$.
The proof that this is the most general re-writing of an arbitrary spin-1 wave function, and is
also equivalent to the popular SB representation, is given in the Supplementary Information (SI). A
similar observation was made in a recent paper by Yukawa and Ueda~\cite{yukawa-ueda}. Average
magnetization depends on $\delta$ as $(\Psi^\dag_{F=1} \v F \Psi_{F=1})^2 /\rho^2 = 1- [\cos
\delta/2]^4$.

The new expression (\ref{eq:our-Psi}) is physically appealing in
that only one new parameter $\delta$, referring to the amount of
mixing between AFM ($\delta=0$) and FM ($\delta=\pi$) components,
appears. Dynamics of spin-1 condensate may be re-examined in light
of the new parametrization scheme. Most importantly, it allows us to
see clearly why dynamics within a specific sub-manifold is
\textit{inherently flawed}, in contrast to recent claims of studying
hydrodynamics of spin-1 condensate within the FM
manifold~\cite{yukawa-ueda,lamacraft}.

Write the $F=1$ spinor wave function confined to the FM sector,
$\Psi_\mathrm{F} = f_\mathrm{F} \bm \eta_F$, with arbitrary complex
scalar function $f_\mathrm{F}$. As is the common practice insert
$\Psi_\mathrm{F}$ into the governing GP equation $i\hbar\partial_t
\Psi = -(\hbar^2/2m)\bm \nabla^2 \Psi$ and take projections on the
left with all independent spinors. For spin-1/2 condensates there
are two such spinors, commonly denoted $\v z$ (a CP$^1$ field) and
its time-reversal counterpart $\v z^t$. The two projections yield
various hydrodynamic relations such as mass continuity, Euler
equation, and Landau-Lifshitz equation for
magnetization~\cite{refael,xu-han-hydro}. For the spin-1 condensate,
one realizes there are three independent spinors with which one can
make a projection: $\bm \eta_A$ and $\bm \eta_F$ as already introduced, and a third one, $\bm
\eta_\mathrm{\overline{F}}$, obtained as the Euler rotation of $\bpm
0 & 0 & 1 \epm^T$. Among them $\bm \eta_F$ and $\bm
\eta_{\overline{F}}$ are a time-reversal pair while $\bm \eta_A$ is
its own time-reversal partner. As will be shown, projection with
$\bm \eta_F$ and $\bm \eta_A$ yield the familiar hydrodynamic
equations similar to the ones already derived in the
literature~\cite{yukawa-ueda,lamacraft}. It is the third projection
with $\bm \eta_{\overline{F}}$ that produces some unusual relations,
and eventually destroys the chance for well-defined dynamics within
${\cal M}_\mathrm{AFM}$ or ${\cal M}_\mathrm{FM}$.

The projection with $\bm \eta^\dag_{\overline{F}}$ in the FM sector,
$i\hbar \bm \eta_{\overline{F}}^\dag
\partial_t \Psi_\mathrm{F} = -(\hbar^2/2m)\bm \eta_{\overline{F}}^\dag \bm \nabla^2
\Psi_\mathrm{F}$, gives the result

\ba (\sin \beta \bm \nabla \alpha - i \bm \nabla \beta )^2 =0
\label{eq:FM-paradox}\ea
which is satisfied if and only if $\bm \nabla \alpha = 0 = \bm
\nabla \beta$.  Recall that $\alpha$ and $\beta$ are the two angles
of the vector $\v d=(\sin \beta \cos \alpha, \sin \beta \sin \alpha,
\cos \beta)$ representing the magnetization direction $\bm
\eta_F^\dag \v F \bm \eta_F = \v d$. Our short analysis
unambiguously shows that the magnetization in the FM phase cannot
fluctuate over space, nor in time as examination of other
hydrodynamic relations would reveal. Inclusion of the confining trap
or the interaction fails to alter the conclusion as shown in SI. A
similar problem of over-constraint arises for purely AFM condensate
$\Psi_\mathrm{A} = f_\mathrm{A} \bm \eta_A$ where one can conclude
(see SI)

\ba \partial_\mu [ \rho \v d \times \partial_\mu \v d] = 0, ~~
\left[\partial_t + {\hbar \over m} ( \bm \nabla \theta \cdot \bm
\nabla ) \right] \v d = 0 \label{eq:AFM-paradox}\ea
must hold in addition to other hydrodynamic equations of mass continuity and Landau-Lifshitz
dynamics. Assuming the uniform density, the first equation implies the initial $\v d$-vector
realizing a uniform vector chirality can still maintain dynamics entirely within the AFM manifold.
However, this is only a very special kind of state to prepare (and non-topological at that) while
all other types of initial states (including all possible topological configurations) will be
forced to leave the initial AFM manifold upon time evolution.

The full GP equation, including the trapping potential and interactions (characterized by two
parameters $g_0$ and $g_2$ for the density- and spin-dependent parts~\cite{ho,machida}), projected
onto $\bm \eta_A, \bm \eta_F, \bm \eta_{\overline{F}}$, respectively, are displayed below (For full
details of derivation see SI):

\begin{widetext}
\ba i \hbar \Bigl[ {\cal D}_{A,t} z_\mathrm{A}  + {1\over \sqrt 2}z_F (\v
e_{+} \cdot \partial_t \v d ) \Bigr] &=& {\hbar^2 \over 2m} \Bigl[
-\bigl({\cal D}_{A,\mu}\bigr)^2 z_A  +z_A (\partial_\mu \v d)^2
-\sqrt{2}( {\cal D}_{F,\mu} z_F )( \v e_{+} \cdot \partial_\mu \v d )
-{1\over \sqrt{2}}z_F ( \v e_{+} \cdot \partial^2_\mu \v d )
\Bigr]\nn& & +{1\over 2} m \omega^2 r^2 z_A + g_0 \rho z_A + g_2
\rho z_A z_F^2, \nn
i \hbar \Bigl[ {\cal D}_{F,t}z_F  - {1 \over \sqrt{2}} z_A ( \v
e_{-} \cdot \partial_t \v d ) \Bigr] &=& {\hbar^2 \over 2m}
\Bigl[-\bigl({\cal D}_{F,\mu}\bigr)^2 z_F +{1\over 2} z_F
(\partial_\mu \v d)^2 +\sqrt{2}( {\cal D}_{A,\mu}z_1 )( \v e_{-}
\cdot \partial_\mu \v d ) + {1 \over \sqrt 2} z_A ( \v e_{-} \cdot
\partial^2_\mu \v d ) \Bigr]\nn & & +{1\over 2} m \omega^2 r^2 z_F
+g_0 \rho z_F + g_2 \rho (z_A^2z_F+z_F^3), \nn
-{i \hbar \over \sqrt 2} z_A ( \v e_{+} \cdot \partial_t \v d ) &=&
{\hbar^2 \over 2m }\Bigl[\sqrt{2} ( {\cal D}_{A,\mu}z_1 )(\v e_{+}
\cdot \partial_\mu \v d )  + {1 \over \sqrt 2}z_A  ( \v e_{+} \cdot
\partial^2_\mu \v d ) + {1\over 2}z_F \Bigl(\v e_{+} \cdot
\partial_\mu \v d \Bigr)^2 \Bigr] \nn
&& - g_2 \rho z_F z_A^2. \label{eq:our-hydro-eq} \ea
\end{widetext}
Here, ${\cal D}_{A,\mu} \! \equiv \! \partial_\mu \!+\! i \partial_\mu \theta  \!+\!
{1\over 2}\partial_\mu\ln\rho$ and ${\cal D}_{F,\mu} \! \equiv \!
\partial_\mu \!+\! i \partial_\mu \theta \!+\! i a_{\mu}+ {1\over
2}\partial_\mu\ln\rho$ ($a_\mu \!=\! - \cos \beta \partial_\mu \alpha
\! - \! \partial_\mu \gamma$) are the covariant derivatives appropriate for
AFM and FM manifold, respectively. Repeated space index $\mu$ are
summed above. A triad of orthogonal unit vectors $\v e_x, \v e_y$
($\v e_\pm = \v e_x \pm i \v e_y$) and $\v d$ are introduced above,
as the Euler rotation ${\cal R}(\alpha, \beta, \gamma)=e^{- i \alpha
S_z} e^{- i \beta S_y} e^{- i \gamma S_z}$ acting on the three basis
vectors $(1~  0 ~ 0)^T$, $( 0~ 1~ 0)^T$, and $( 0~ 0~ 1)^T$,
respectively. Spin representation $(S_\alpha )_{\beta \gamma} = -i
\varepsilon_{\alpha\beta\gamma}$ used to construct the rotation
matrix ${\cal R}$ above is distinct from $\v F$ used earlier to
construct another Euler rotation matrix ${\cal U}$. Anomalies
mentioned earlier within the AFM or the FM phase can be derived as
limits of Eq. (\ref{eq:our-hydro-eq}) with $(z_A , z_F ) = (1,0)$ or
$(z_A , z_F ) = (0,1)$.
\\

{\it Physical implications}.- An immediate conclusion of our analysis is that temporal
dynamics whose initial state is defined solely within ${\cal M}_\mathrm{FM}$ or ${\cal
M}_\mathrm{FM}$ should instantly evolve into a mixed phase containing both components
with a certain mixing angle $\delta$. Indeed, examination of the GP dynamical equation
reveals this to be the case. But first, let us show that even the familiar results
addressing small fluctuations around ground states from the early literature on spinor
dynamics can, and must, be interpreted within the framework we propose~\cite{ho,machida}.
Assuming infinite medium and neglecting the density fluctuation which is massive, states
around the fully polarized FM ground state with its magnetization along the $(0,0,1)$
direction is captured by the wave function of the form, Eq. (\ref{eq:our-Psi}):

\ba \Psi_{\mathrm{FM}+\delta\cdot \mathrm{AFM}} = \sqrt{\rho} e^{i \theta}
\left[   \bpm 1  \\ {1 \over \sqrt{2} }\beta e^{i \alpha}  \\
0 \epm + \delta \bpm 0   \\ 1 \\
0 \epm  \right] .  \ea
Standard spin-wave dispersion $\hbar \omega_{\v k} = \hbar^2 \v k^2 /2m$ follows from feeding the above
wave function into the GP equation as shown in SI. States around the AFM-ordered state is
described, in turn, by the wave function

\ba \Psi_{\mathrm{AFM}+\delta\cdot\mathrm{FM}} = \sqrt{\rho} e^{i \theta} \left[ \bpm -{1 \over \sqrt{2}}d_-   \\ 1 \\
{1 \over \sqrt{2}}d_+ \epm  + \delta \bpm  e^{-i (\alpha + \gamma)}
\\ 0 \\ 0 \epm \right], \label{eq:AFM-fluctuated-wf}\ea
$d_\pm =d_x \pm i d_y =  \beta e^{\pm i \alpha}$,  with the dispersion $\hbar \omega_{\v k} =
\sqrt{{\hbar ^2 \v k^2\over 2 m} \left({\hbar^2 \v k^2\over 2m } + 2 g_2 \rho \right)}$ consistent
with the result of Refs. \cite{ho,machida}. These exercises already point out that temporal
dynamics around a fully FM or a fully AFM state must involve fluctuations into the ``other
manifold".

Now let's turn to the fate of topological objects in light of our new formulation of temporal
dynamics. There are various vortex configurations in both AFM and FM manifolds which realize a
non-trivial $\pi_1$-homotopy~\cite{ueda-review}. A recent study clearly showed that the putative
manifold of ${\cal M}_\mathrm{AFM}$ or ${\cal M}_\mathrm{FM}$, on which the topology consideration
initially rests, must give way in favor of the order parameter $\Psi$ rotating into the larger
manifold $S^5$ in the core region in order to avoid the catastrophe of kinetic energy
divergence~\cite{ueda-vortex}. It implies that even the static, equilibrium configuration, let
alone the dynamics, cannot be properly defined for vortex object within the restricted manifold.
Meanwhile our numerical solution of the dynamic GP equation readily showed that the initial vortex
profile in the far region at time $t=0$, \textit{which can be defined entirely within} ${\cal
M}_\mathrm{AFM}$ {\it or} ${\cal M}_\mathrm{FM}$, immediately evolves into a mixed phase upon
$t>0$.

The emergence of mixing in the temporal evolution of topological objects is more dramatically seen
for the Skyrmion - a realization of non-trivial $\pi_2$-homotopy in two dimensions. While ${\cal
M}_\mathrm{FM}=\mathrm{SO}(3)$ does not support a two-dimensional topological defect due to trivial
second homotopy $\pi_2 (\mathrm{SO(3)} ) = 0$, the antiferromagnetic manifold does allow one due to
$\pi_2 ({\cal M}_\mathrm{AFM}) = \mathbb{Z}$. Free of the kinetic energy divergence, it is possible
to construct a wave function for the Skyrmion entirely within the AFM
manifold~\cite{ueda-review,ueda-vortex}:

\ba \Psi_{A}^{(S)} (\v r) =\sqrt{\rho(r)} \eta_A [\v d_S ]. \ea
The $\v d_S$-vector field realizes a Skyrmion with finite Skyrmion number $(1/4\pi)\int \v d_S
\cdot (\partial_x \v d_S \times \partial_y \v d_S ) =1$ (Explicit form of $\v d_S$ is given in SI).
Feeding such a configuration with suitably optimized density profile $\rho(r)$ as an initial state
in the GP equation, the temporal dynamics was solved to reveal the fate of the topological Skyrmion
state.

\begin{figure*}[t]
\centering
\includegraphics[width=1\textwidth]{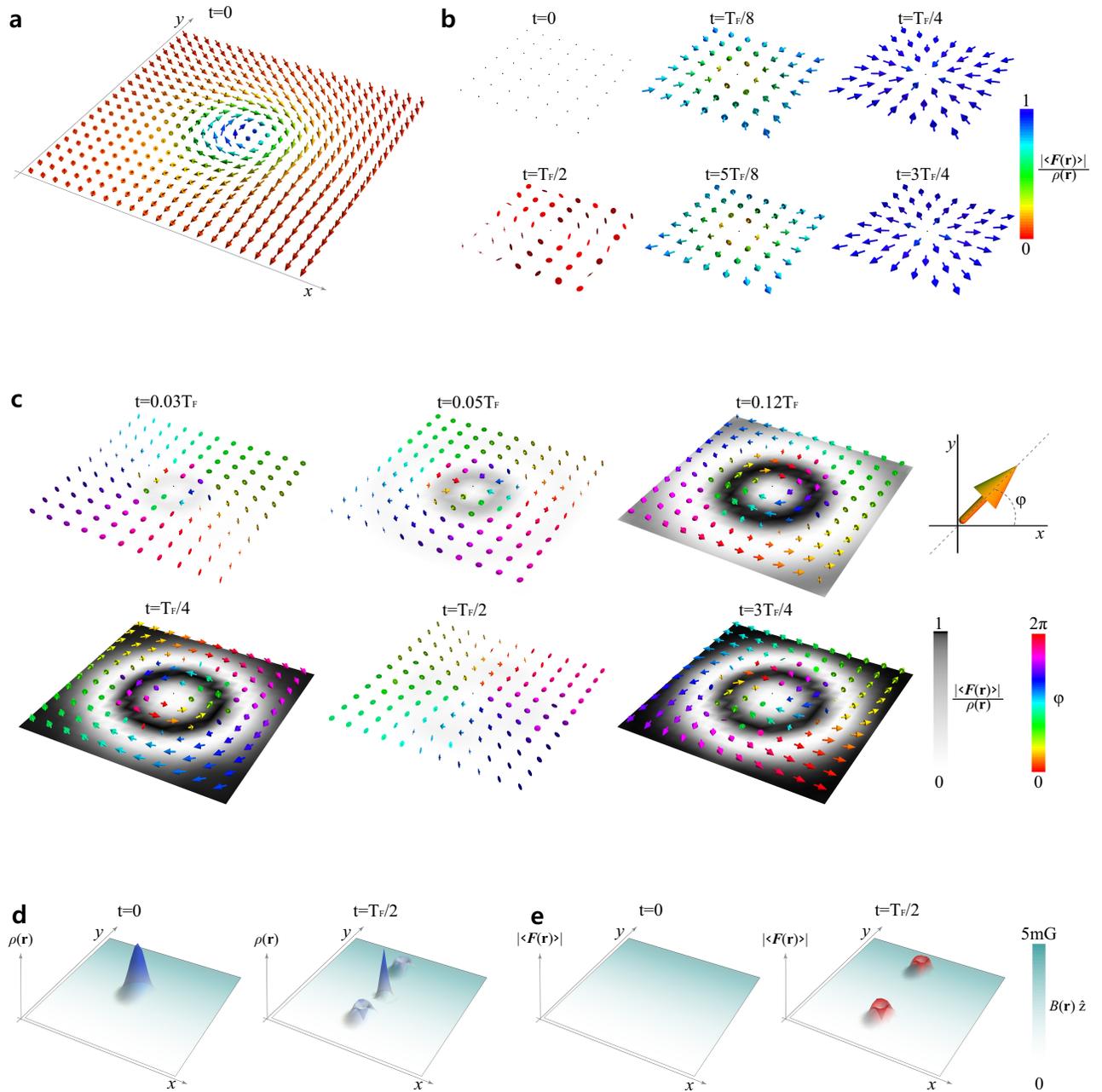}
\caption{(a) Initial Skyrmion configuration for $\v d$-vector prepared in the AFM condensate. (b)
Temporal evolution of the (normalized) magnetization pattern $\Psi^\dag \v F \Psi/\rho$.
Well-defined magnetizations oscillate with period $T_\text{F} \simeq 6.5 \,\omega^{-1}$.
Dimensionless interaction parameters for the calculation are $( g_0 , g_2 )N/\hbar \omega = (100,
10)$ (See SI for details of numerical solution). (c) Temporal evolution of the magnetization under
uniform magnetic field $\v B = B_0 \hat{z}$, $B_0 = 1.6$mG~\cite{berkeley}. Regions of strong
magnetization form a ring-like pattern. Period of magnetization oscillation is  $T_\text{F} \simeq
6.7 \,\omega^{-1}$ ($\omega$=frequency of the trapping potential). (d),(e) Time evolution under
magnetic field gradient $\v B = B(y) \hat{z}$, $dB(y)/dy = 1.6\times 10^{-4}$ G/$\mu$m. Width of
the initial gaussian density was $\sigma = 2.2\mu$m. Well-defined pockets of non-zero magnetization
separate out from the central, AFM region due to the field gradient, then oscillate back and forth
along the $y$-direction with period $T_\text{F} \simeq 6.5 \,\omega^{-1}$. }
\label{fig:2d_dynamics}
\end{figure*}

As described in Fig. \ref{fig:2d_dynamics}, the most striking feature of the simulation is the
immediate appearance of non-zero magnetization $\Psi^\dag (\v r , t) \v F \Psi (\v r , t)$ through
the entire condensate for $t>0$.  How these magnetization patterns evolve over time for zero,
uniform, and non-uniform magnetic fields were studied numerically. In all cases magnetization
dynamics we have revealed are quite intricate. For instance a ring of strongly magnetized region is
formed around the Skyrmion center by the uniform magnetic field (Fig. 1c). The magnetic field
gradient creates satellites of magnetized condensates that move periodically apart from the central
AFM region (Fig. 1d,e). Whether spontaneous or induced by magnetic field, the formation of
magnetized regions signals the breakdown of the initial AFM manifold at all times of the temporal
evolution. In short, Skyrmion is no longer a well-defined topological entity. Previous theoretical
studies were focused on the static configuration of the Skyrmion~\cite{ueda-review} without careful
consideration of how the AFM manifold will sustain itself over time.

Although it may seem disastrous at first that the nice topological notion associated with various
sub-manifolds of spin-1 condensate become ill-defined dynamically, it also might open up
interesting possibilities from the engineering perspective of manipulating the condensate. The
separation of FM components out of the initial AFM condensate by the magnetic field gradient
discussed above offers one such example. The key to condensate engineering is to prepare the
condensate wholly in the AFM (or FM) condensate with a textured $\v d$-vector. According to Eqs.
(\ref{eq:FM-paradox}) and (\ref{eq:AFM-paradox}) it is impossible for such initial $\v d$-vector to
evolve within the same manifold and must induce the other component immediately, opening up
interesting potential for realizing various mixed phases.  We demonstrate this point with another
example.

\begin{figure*}[t]
\centering
\includegraphics[width=1\textwidth]{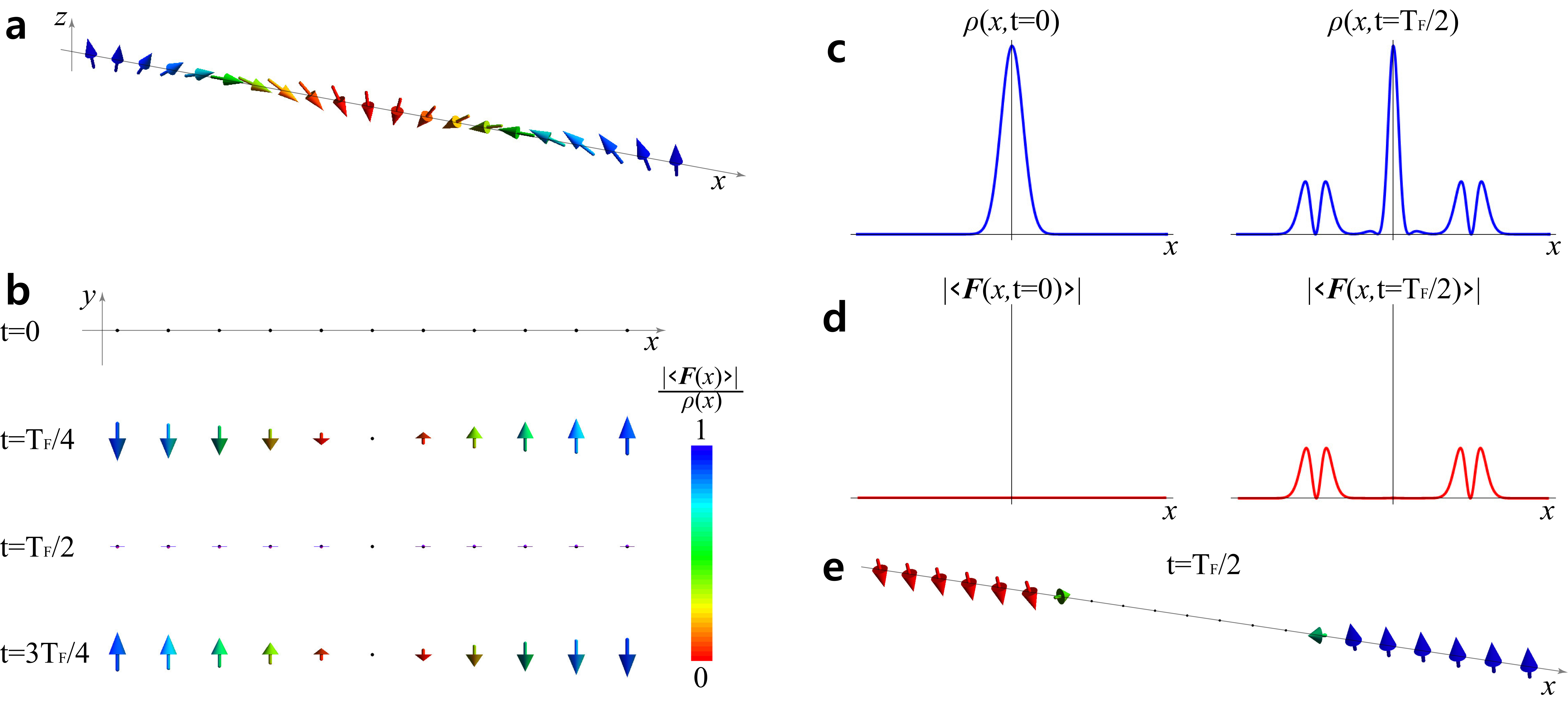}
\caption{ (a) Initial $\v d$-vector texture in the one-dimensional AFM condensate. (b) Temporal
evolution of the (normalized) magnetization $\Psi^\dag \v F \Psi/\rho$. Well-defined magnetizations
oscillate with period $T_\text{F} \simeq 6.5 \,\omega^{-1}$. The central region effectively becomes
the domain wall between magnetizations pointing along $+\hat{y}$ and $-\hat{y}$. Dimensionless
interaction parameters are $( g_0 , g_2 )N/\hbar \omega  = (100, 10)$. (c),(d) Temporal evolution
of the density $\rho(x) = \Psi^\dag \Psi$ and the magnetization $\langle \v F (x) \rangle =
\Psi^\dag \v F \Psi$ under magnetic field gradient $\v B = B(x) \hat{z}$,  $dB(x) /dx = 2.0 \times
10^{-4}$ G/$\mu$m. Initial gaussian packet has the width $\sigma = 2.0 \mu$m. Well-defined
magnetized satellites develop on both sides of the central AFM peak, then oscillate back and forth
along the $x$-direction with period $T_\text{F} \simeq 6.5 \,\omega^{-1}$. (e) Corresponding
(normalized) magnetization pattern at time $t= T_F/2$. Arrows (magnetization vectors) are aligned
with the local field direction.} \label{fig:1d_dynamics}
\end{figure*}

Take a one-dimensional, non-topological but non-uniform AFM
initial state with a twisted $\v d$-vector

\ba \Psi_{A}^{(1D)} (x, t=0) = {\sqrt{\rho(x)} \over x^2\!+\! R^2} \bpm -\sqrt{2} R x \\ x^2 \!-\! R^2
\\ \sqrt{2} R x \epm, \label{eq:1d_AFM} \ea
where the density $\rho(x) \sim \exp (-x^2 /2\sigma^2)$ is a gaussian of width $\sigma$. The $\v
d$-vector itself executes a cycloidal rotation within the $xz$ plane. Subject to temporal
evolution, this initial configuration develops a pattern of magnetization at later times $t>0$ in
the $y$-direction as shown in Fig. \ref{fig:1d_dynamics}. Amusingly, even a domain wall forms
between regions of magnetizations $+\hat{y}$ and $-\hat{y}$ (Fig. \ref{fig:1d_dynamics}b). Applying
the field gradient causes magnetized satellite peaks to emerge in similarity to the two-dimensional
Skyrmion case.

Our work suggests a new and intuitive way to view the $F=1$ condensate dynamics in terms of a
mixture of AFM and FM base manifolds. The dynamical equations are well-defined only by mixing the
AFM and FM components at all times and space. The notion of topology previously regarded as valid
in AFM or FM sub-manifold is therefore to be discarded. In its place, the easy and inevitable
mixing of the manifolds opens up new possibilities for engineering the topology of the $F=1$
condensate by means of well-designed magnetic fields.

\acknowledgments{This work is supported by the NRF grant (No. 2013R1A2A1A01006430). P. J.
K. acknowledges support from the Global Ph. D. Fellowship Program (NRF-2012). J. H. H. is
indebted to professor Hui Zhai for inspiring him into the field of spinor BEC as well as
for his guidance, and to professor Yong-il Shin for discussion of his experiment.}

\begin{thebibliography}{24}

\bibitem{ueda-review} Y. Kawaguchi and M. Ueda, Phys. Rep. \textbf{520}, 253 (2012).

\bibitem{khalat} I. M. Khalatnikov,
\textit{An Introduction to the Theory of Superfluidity}, translated
by P. C. Hohenberg (W. A. Benjamin Inc., New York, 1965).

\bibitem{lamacraft} A. Lamacraft, Phys. Rev. A \textbf{77},
063622 (2008).

\bibitem{refael} B. Barnett, D. Podolsky, and G. Refael,
Phys. Rev. B \textbf{80}, 024420 (2009).

\bibitem{yukawa-ueda} E. Yukawa and M. Ueda, Phys. Rev. A \textbf{86}, 063614 (2012).

\bibitem{ho} T.-L. Ho, Phys. Rev. Lett. \textbf{81}, 742 (1998).

\bibitem{zhou} F. Zhou, Phys. Rev. Lett. \textbf{87}, 080401 (2001).


\bibitem{shin} J. Choi, W. J. Kwon, and Y. Shin, Phys. Rev. Lett. \textbf{108}, 035301 (2012).

\bibitem{xu-han} X.-Q Xu and Jung Hoon  Han, Phys. Rev. A \textbf{86}, 063619
(2012).

\bibitem{ueda-vortex} S. Kobayashi, Y. Kawaguchi, M. Nitta, and M. Ueda, Phys. Rev. A
\textbf{86}, 023612 (2012).

\bibitem{xu-han-hydro} X.-Q. Xu and J. H. Han, Phys. Rev. Lett. \textbf{108}, 185301 (2012).

\bibitem{machida} T. Ohmi and K. Machida, J. Phys. Soc. Jpn. \textbf{67}, 1822 (1998).

\bibitem{berkeley} L. E. Sadler, J. M. Higbie, S. R. Leslie, M. Vengalattore, and D. M. Stamper-Kurn,
Nature \textbf{443}, 312-315 (2006).

\end{thebibliography}
\end{document}


\title{Fate of Topology in Spin-1 Spinor Bose-Einstein Condensate: Supplementary Information}

\author{Yun-Tak Oh}
\affiliation{Department of Physics, Sungkyunkwan University, Suwon
440-746, Korea}
\author{Panjin Kim}
\affiliation{Department of Physics, Sungkyunkwan University, Suwon
440-746, Korea}
\author{Jin-Hong Park}
\affiliation{Department of Physics, Sungkyunkwan University, Suwon
440-746, Korea}
\author{Jung Hoon Han}
\email[Electronic address:$~~$]{hanjh@skku.edu}
\affiliation{Department of Physics, Sungkyunkwan University, Suwon
440-746, Korea} \affiliation{Asia Pacific Center for Theoretical
Physics, POSTECH, Pohang, Gyeongbuk 790-784, Korea}

\begin{abstract} This note discusses several technical details not covered in the main text of the paper.
\end{abstract}
\pacs{03.75.Kk, 03.75.Mn, 05.30.Jp}\maketitle

\section{Proof of equivalence to Schwinger boson representation of
spin-1 condensate}
In the paper we suggested that an arbitrary $F=1$ spinor BEC wave function may be cast in the form

\ba \Psi_{F=1} =\sqrt{\rho}e^{i\theta} \bm \eta , \ea
where the spinor part $\bm \eta$ is a linear combination of the AFM $(\bm \eta_A)$ and the FM $(\bm
\eta_F)$ parts,

\ba \bm \eta = z_A \bm \eta_A + z_F \bm \eta_F , ~ \bpm z_A \\
z_F \epm = \bpm \cos \delta/2 \\ e^{i\tau} \sin \delta/2 \epm ,
\label{eq:our-decomposition}\ea
with a pair of coefficients $(z_A , z_F )$ obeying the constraint $|z_A |^2 + |z_F |^2 = 1$. Due to
the global phase $e^{i\theta}$ entering in the wave function $\Psi_{F=1}$ one can freely choose
$z_A$ to be real without loss of generality. Furthermore, the phase factor $e^{i\tau}$ of $z_F$
always appears multiplied by $e^{-i\gamma}$ of $\bm \eta_F$ and can be absorbed by it:
$e^{i(\tau-\gamma)} \rightarrow e^{-i\gamma}$. This leads to the simplified expression

\ba \Psi_{F=1}= \sqrt{\rho} e^{i\theta} \left[\bm \eta_A \cos {\delta \over 2} + \bm \eta_F \sin {\delta
\over 2} \right] \label{eq:our-decomposition} \ea
given in Eq. (2) of the paper with

\ba \bm \eta_A\!=\! \bpm -{1\over\sqrt{2}} e^{-i\alpha} \sin \beta
\\ \cos\beta \\ {1\over\sqrt{2}} e^{i\alpha} \sin \beta \epm, ~
\bm \eta_F \!=\! e^{-i\gamma}
\bpm e^{-i\alpha} \cos^2 {\beta \over 2 } \\
{1\over\sqrt{2}} \sin \beta \\ e^{i\alpha} \sin^2 {\beta \over 2 }
\epm . \label{eq:eta1-eta2} \nn \ea

One can furthermore prove that the new scheme is equivalent to the standard Schwinger boson (SB)
expression of the (unnormalized) spin-1 wave function
[\onlinecite{spin-half-hydro-refeal,ueda-schwinger}]

\ba
\Psi_\mathrm{SB} \sim \sqrt{\rho} e^{i\theta_\mathrm{SB}} (u_1 a^\dag + v_1 b^\dag ) (u_2
a^\dag + v_2 b^\dag ) |0\rangle .\ea
Each $(u_i, v_i ) = (\cos \theta_i /2 , e^{i \phi_i } \sin \theta_i /2 )$ defines a point $\v n_i$
on the unit sphere S$^2$ through the CP$^1$ mapping. The FM state is obtained by identifying
$\theta_1 = \theta_2$, $\phi_1 = \phi_2$, or $\v n_1 = \v n_2$. The antiferromagnetic phase is
identified with $\v n_2 = -\v n_1$, by writing $(u_2, v_2 ) = (v^*_1, - u_1 )$. Keeping in mind
that the angular momentum state $|J, m\rangle$ in the SB representation is

\ba |J, m\rangle = {1\over\sqrt{(J\!+\!m)!(J\!-\!m)!}} (a^\dag)^{J\!+\! m} (b^\dag)^{J\!-\!m}
|0\rangle , \ea
the SB wave function becomes, together with proper normalization factor,

\ba \Psi_\mathrm{SB} = \sqrt{\rho} e^{i\theta_\mathrm{SB}} \sqrt{2 \over 3+ \v n_1 \cdot \v n_2}
\bpm \sqrt{2} u_1 u_2 \\ u_1 v_2 +u_2 v_1 \\ \sqrt{2} v_1 v_2 \epm. \label{eq:SB-form}\ea

Components of the original spinor wave function $\Psi_{F=1}$ are related to the SB decomposition
through the roots of the quadratic equation

\ba \psi_+ z^2 + \sqrt{2} \psi_0 z + \psi_- =0 , \label{eq:for-z} \ea
with the coefficients derived from the wave function $\Psi_{F=1} = ( \psi_+ , \psi_0, \psi_- )^T$.
The two roots ought to correspond, precisely, to $z_1 = - v_1/ u_1 $ and $z_2=- v_2 / u_2 $ if Eq.
(\ref{eq:SB-form}) should hold for arbitrary $F=1$ wave function~\cite{ueda-schwinger}. Once the
roots of a particular $F=1$ wave function are found they can be related to the SB parameters
through the formula

\ba z_i = -{v_i \over u_i}= -e^{i \phi_i}\tan{\theta_i \over 2} .\label{z sol1} \ea

We may now apply this procedure to the new condensate wave function $\Psi_{F=1}$ shown in Eq.
(\ref{eq:our-decomposition}). The two roots are readily found to be

\ba z_1 &=& - e^{i \alpha} \tan{\beta \over 2}, \nn
z_2 &=& e^{i
\alpha}\left(\cot{\beta \over 2}\!+\! {1 \over \sqrt{2} e^{i \gamma}
\cot {\delta \over 2} \sin^2 {\beta \over 2} \!-\! {\sin \beta \over 2} }
\right). \nn \label{eq:z1-z2} \ea
From the first solution it follows that the first pair of SB parameters $(\phi_1, \theta_1)$ is
equal to $(\alpha, \beta)$ in the wave function $\Psi_{F=1}$. That is, once a SB construction is
given, the corresponding $\Psi_{F=1}$ can be found by identifying $(\alpha, \beta)$ with the first
pair of SB parameters $(\phi_1, \theta_1 )$. The second solution $z_2$ can then be used to relate
the remaining unknowns $(\gamma, \delta)$ to the SB parameters.

Rather than trying to tackle the second of Eq. (\ref{eq:z1-z2}) directly, we adopt a different, more pragmatic way to relate $(\gamma, \delta)$ in terms of SB parameters. It proves quite useful to compare the magnetization averages for each representation. From $\Psi_\mathrm{SB}$ one finds

\ba \v S_\mathrm{SB} = \Psi^\dag_\mathrm{SB} \v F \Psi_\mathrm{SB}  = \rho {2(\v n_1 + \v n_2
)\over 3 + \v n_1 \cdot \v n_2}.\ea
The average from $\Psi_{F=1}$ is more complicated, but fortunately one can relate it to the Euler rotation of some basis spinor

\ba \v S_{F=1} = \Psi^\dag_{F=1}\v F \Psi_{F=1} =  \rho {\cal R}(\alpha, \beta, \gamma) \bpm  {1 \over \sqrt 2}\sin \delta \\  0 \\
 \sin^2 {\delta \over 2} \epm ,\nn \label{eq:R}\ea
with ${\cal R}(\alpha, \beta, \gamma) = e^{- i \alpha S_z} e^{- i \beta S_y} e^{- i \gamma S_z}$ and $[S_\alpha ]_{\beta\gamma}
= -i \varepsilon_{\alpha\beta\gamma}$.

Squaring each average and identifying the two, $\v S_\mathrm{SB}^2 = \v S_{F=1}^2$, gives out the relation

\ba 1- \left( \cos {\delta \over 2} \right)^4 = 8{\bigl(1+\v n_1 \cdot \v n_2
\bigr) \over \bigl(3+ \v n_1 \cdot \v n_2 \bigr)^2 } .\ea
The mixing angle $\delta$ is obtained easily as

\ba \cos {\delta
\over 2}= \sqrt{ {1-\v n_1 \cdot \v n_2 \over 3+ \v n_1 \cdot \v
n_2}}. \label{eq:for-delta}\ea
For any SB wave function one can first work out $\v n_1$ and $\v n_2$, then use the above relation
to uniquely specify $\delta$ in Eq. (\ref{eq:our-decomposition}). Proper limits are recovered when
$\v n_1 = \v n_2$ (FM, $\delta = \pi$) and $\v n_1 = -\v n_2$ (AFM, $\delta = 0$).

To determine the remaining parameter $\gamma$, one notes that $\v S_{F=1}$ given in Eq.
(\ref{eq:R}) ought to be orthogonal to the following vector:

\ba{\cal R}(\alpha,\beta,\gamma)\bpm 0\\1\\0 \epm  =\bpm-\cos\alpha\cos\beta\sin\gamma -\sin\alpha \cos\gamma\\
-\sin\alpha\cos\beta\sin\gamma+ \cos\alpha\cos\gamma \\
\sin\beta\sin\gamma \epm . \nn \label{eq:1.15}\ea
Taking the inner product of $\v S_{F=1}$, or equivalently of $\v S_\mathrm{SB}$, with Eq.
(\ref{eq:1.15}) should give zero:

\ba \sin\gamma
\bigl(\sin\beta\cos\theta_2-\cos(\alpha-\phi_2)\cos\beta
\sin\theta_2 \bigr)\nn - \cos\gamma\sin(\alpha-\phi_2)\sin\theta_2 =
0. \ea
Luckily this equation contains $\gamma$ only, and one finds

\ba \tan \gamma =
{\sin\theta_2\sin(\alpha- \phi_2) \over \sin\beta\cos\theta_2
-\cos(\alpha - \phi_2)\cos\beta \sin\theta_2 }.\nn \label{eq:for-gamma}\ea
All the variables on the right side are the SB parameters, already assumed known or given in advance.

Once a particular SB parametrization of the $F=1$ wave function is given, one can proceed
systematically to find the corresponding $(\alpha, \beta, \gamma, \delta)$ of the new decomposition
scheme, Eq. (\ref{eq:our-decomposition}), by (i) identifying $(\alpha, \beta ) = (\phi_1 , \theta_1
)$, and (ii) solving for $(\gamma, \delta)$ using Eqs. (\ref{eq:for-delta}) and
(\ref{eq:for-gamma}). Finally, the global phase factor $\theta$ in $\Psi$ is found by taking the
projection

\ba e^{i \theta } = \bm \eta^\dag \Psi_\mathrm{SB} \ea
now that $\bm \eta$ is completely fixed. With the one-to-one correspondence to the SB wave function
established, Eq. (\ref{eq:our-decomposition}) constitutes a new, alternative way to express the
most general spin-1 wave function.

\section{Decomposition of the Gross-Pitaevskii equation}
In this section the Gross-Pitaevskii (GP) equation for spin-1 condensate as shown in Eq. (5) of
the paper is derived. The standard form of the GP equation,

\ba i\hbar {\partial \over \partial t}\Psi = -{\hbar^2 \over 2m} \bm
\nabla^2 \Psi+ {1 \over 2} m \omega^2 r^2 \Psi + g_0 \rho \Psi + g_2 \v S \cdot \v F \Psi , \label{GP eq}\nn\ea
will be re-analyzed based on the decomposition being proposed in Eq. (\ref{eq:our-decomposition}).
On direct insertion of $\Psi_{F=1}$ into the GP equation one finds

\begin{widetext}

\ba && i\hbar  \Bigl  ([\partial_t \omega_A +i \omega_A \partial_t \theta ] \bm \eta_A
+ [\partial_t \omega_F + i \omega_F \partial_t \theta  ]\bm\eta_F+ \omega_A[\partial_t
\bm\eta_A]+\omega_F[\partial_t \bm\eta_F]
\Bigr)\nn
&& = {-\hbar^2\over 2m} \Bigl(\bigl[\partial^2_\mu \omega_A + 2 i
\partial_\mu \omega_A \partial_\mu \theta  + \omega_A \bigl(i\partial^2_\mu
\theta- (\partial_\mu \theta)^2\bigr)\bigr]\bm \eta_A+
2\bigl[\partial_\mu \omega_A + i \omega_A \partial_\mu \theta \bigr]
\bigl[\partial_\mu \bm \eta_A \bigr] +\omega_A\bigl[\partial^2_\mu
\bm\eta_A\bigr]\nn
& & ~~~~~~~~~~+\bigl[\partial^2_\mu \omega_F + 2 i\partial_\mu \omega_F \partial_\mu \theta
 + \omega_F \bigl(i\partial^2_\mu
\theta- (\partial_\mu \theta)^2\bigr)\bigr]\bm \eta_F+ 2\bigl[ \partial_\mu \omega_F + i\omega_F
\partial_\mu \theta \bigr] \bigl[\partial_\mu \bm \eta_F \bigr] +\omega_F \bigl[\partial^2_\mu
\bm\eta_F\bigr] \Bigr) \nn
& &+{1 \over 2} m \omega^2 r^2
\bigl[\omega_A \bm \eta_A + \omega_F \bm \eta_F \bigr] + g_0\rho \bigl[
\omega_A \bm \eta_A + \omega_F \bm \eta_F \bigr] +  g_2 \v S \cdot
\v F \bigl[\omega_A \bm\eta_A + \omega_F \bm \eta_F \bigr] .
\label{eq:our-GP}\ea
\end{widetext}
Here $\omega_A$  and $\omega_F$ are temporary abbreviations for $\omega_A \equiv \sqrt{\rho}z_A$
and $\omega_F \equiv \sqrt{\rho}z_F$.
Note that repeated $\mu$ implies summation for $\v r = x,~y,~z $. As in Ref. \onlinecite{refael} we expect to obtain various
hydrodynamic relations by projecting the above equation with $\bm \eta_A$ and  $\bm\eta_F$ already
introduced in Eq. (\ref{eq:eta1-eta2}), and with a third one, $\bm \eta_{\overline{F}}$, defined by

\ba \bm \eta_{\overline{F}} =\bpm e^{- i \alpha} \sin^2 { \beta \over 2} \\
{ 1 \over \sqrt{2}} \sin \beta \\ e^{ i \alpha} \cos^2 { \beta \over 2} \epm .\ea
As discussed in the main text, $\bm \eta_F, \bm \eta_A, \bm \eta_{\overline{F}}$ are the Euler
rotations ${\cal U}(\alpha, \beta, \gamma)$ of the three basis spinors $(1~ 0 ~ 0)^T$, $(0~1~0)^T$
and $(0~0~1)^T$, respectively. Only this time the Euler rotation ${\cal U}(\alpha, \beta, \gamma) =
e^{-i\alpha F_z} e^{-i\beta F_y } e^{-i\gamma F_z }$ is generated with a different set of spin
matrices $\v F$.

It will be seen shortly that projection with the first two spinors yield standard hydrodynamic
relations such as mass continuity, Euler equation, and the Landau-Lifshitz equation. The projection
with the third spinor, however, has been neglected in the past literature\cite{lamacraft} and yield
some surprising consequences.

Before proceeding to the hydrodynamic decomposition, some mathematical preliminaries are in order.
It proves extremely convenient to define a triad of orthogonal basis vectors,

\begin{widetext}
\ba \v e_x &\equiv&  {\cal R} \bpm 1 \\ 0 \\ 0 \epm = ( \cos \beta \cos \alpha \cos\gamma
-\sin\alpha \sin\gamma , \cos \beta \sin \alpha \cos\gamma +\cos\alpha \sin\gamma , -\sin
\beta\cos\gamma ) , \nn
\v e_y &\equiv&  {\cal R} \bpm 0 \\ 1 \\ 0 \epm = ( -\sin \alpha\cos\gamma
-\cos\alpha\cos\beta \sin\gamma  , \cos \alpha\cos\gamma -\cos\beta\sin\alpha \sin\gamma
, \sin\beta \sin \gamma ) , \nn
\v d & \equiv & {\cal R} \bpm 0 \\ 0 \\ 1 \epm =  (\sin \beta \cos \alpha, \sin \beta \sin \alpha,
\cos \beta), \label{eq:ex-ey-d}\ea
\end{widetext}
using the rotation matrix ${\cal R}$ introduced earlier in Eq. (\ref{eq:R}). For later convenience
one also defines $\v e_\pm \equiv \v e_x \pm i \v e_y$. Note that our definitions of the triad $(\v
e_x , \v e_y , \v d)$ are more general than those of Ref. \onlinecite{refael}.

Various connections can be defined among the three spinors: $-i \bm \eta_\alpha^\dag [ \partial_\mu
\bm \eta_\beta ]$. Some are zero, while all the non-zero connections can be related to the
derivatives of the geometric quantities defined in Eq. (\ref{eq:ex-ey-d}):

\begin{widetext}
\ba \v e_x \cdot \partial_\nu \v e_y &=& -\cos\beta \partial_\nu \alpha - \partial_\nu \gamma  = -i
\bm\eta^\dag_F [ \partial_\nu \bm\eta_F ]= i \bm\eta^\dag_{\overline{F}} [ \partial_\nu
\bm\eta_{\overline{F}} ], \nn
\v e_{+} \cdot \partial_\nu \v d &=& e^{-i \gamma } \bigl(\partial_\nu\beta + i \sin\beta \partial_\nu \alpha\bigr) ={\sqrt 2}  \bm\eta^\dag_A [ \partial_\nu \bm\eta_F ] = \sqrt{2} \bm\eta^\dag_{\overline{F}} [ \partial_\nu \bm\eta_A ], \nn
\v e_{-} \cdot \partial_\nu \v d &=&e^{i \gamma }\bigl( \partial_\nu \beta -i \sin\beta \partial_\nu \alpha\bigr)
= -{\sqrt 2}\bm\eta^\dag_F [ \partial_\nu \bm\eta_A ] , \label{geometric relation}
\ea
\end{widetext}
where $\nu$ = $(\v r, t)$ .
The connection $-i\bm \eta^\dag_F \partial_\nu \bm \eta_F$ appears frequently in the hydrodynamic
equations and will be labeled $a_\nu$:

\ba a_{\nu} = -\cos\beta \partial_\nu \alpha - \partial_\nu \gamma.
\ea
Crucial to the hydrodynamic formulation are the various projections of the second derivatives of
the spinor, $\bm \eta_\alpha^\dag \partial^2_\mu \bm \eta_\beta$, which can be re-written nicely in
terms of geometric quantities:

\ba \bm\eta^\dag_A
\partial^2_\mu \bm\eta_A&=&  -\bigl(\partial_\mu \v d\bigr)^2 , \nn
\bm\eta^\dag_A \partial^2_\mu \bm\eta_F &=& i\sqrt{2} a_{\mu} ( \v e_+ \cdot \partial_\mu \v d ) +
{1 \over \sqrt 2} \v e_+ \cdot \partial^2_\mu \v d , \nn
\bm\eta^\dag_F \partial^2_\mu \bm\eta_A&=&  -{1 \over \sqrt 2} \v e_+ \cdot \partial^2_\mu \v d , \nn
\bm\eta^\dag_F \partial^2_\mu \bm\eta_F&=& i \partial_\mu a_{\mu} - \Bigl({1 \over 2}\bigl(\partial_\mu \v d\bigr)^2 +\bigl(a_{\mu}\bigr)^2 \Bigr) , \nn
\bm\eta^\dag_{\overline{F}}\partial^2_\mu \bm\eta_A&=& {1 \over \sqrt 2} \v e_+ \cdot \partial_\mu^2 \v d , \nn
\bm\eta^\dag_{\overline{F}} \partial^2_\mu \bm\eta_F&=& {1 \over 2} \Bigl(\v e_+ \cdot \partial_\mu \v d \Bigr)^2.
\ea
For convenience one can also define partial magnetization $ \bm \eta_\alpha^\dag \v F \bm
\eta_\beta = \v S_{\alpha\beta}$, which gives

\ba \v S_{AF} = \v S_{\overline{F}A} &=& {1 \over \sqrt{2}} (\v e_x + i \v e_y ), \nn
\v S_{FA} = \v S_{A\overline{F}} &=& {1 \over \sqrt{2}} (\v e_x - i \v e_y ), \nn
\v S_{FF} &=&  \v d .\ea
Magnetization for the general wave function $\Psi_{F=1}$ becomes $ \v S = \rho \bigl( \sqrt{2} z_A
z_F \v e_x + z_F^2 \v d\bigr)$. Other useful relations are $\v S_{AA}=\v S_{\overline{F}F}=\v
S_{F\overline{F}} = 0$, and $\v S_{\alpha\beta} \cdot \v d = \delta_{F\alpha}\delta_{F \beta}$.

After these technical preparations we can start to decompose the
equation (\ref{eq:our-GP}) by projecting first with $\eta^\dag_A$:

\begin{widetext}

\ba i \hbar \Bigl[ {\cal D}_{A,t} z_A  + {1\over \sqrt 2}z_F \bigl( \v e_{+} \cdot \partial_t \v d
\bigr) \Bigr] &=& {\hbar^2 \over 2m} \Bigl[ -\bigl({\cal D}_{A,\mu}\bigr)^2 z_A  +z_A (\partial_\mu
\v d)^2 -\sqrt{2} ( {\cal D}_{F,\mu} z_F )\bigl( \v e_{+} \cdot \partial_\mu \v d \bigr)  -{1\over
\sqrt{2}}z_F \bigl( \v e_{+} \cdot
\partial^2_\mu \v d \bigr ) \Bigr]  \nn
& & +{1\over 2} m \omega^2 r^2 z_A + g_0 \rho z_A + g_2 \rho z_A z_F^2 . \label{geometric eta1 products}
\ea
Inner product with $\eta^\dag_F$ gives

\ba   i \hbar \Bigl[{\cal D}_{F,t}z_F  - {1 \over \sqrt{2}} z_A \bigl (\v e_{-} \cdot \partial_t \v
d \bigr)  \Bigr]&=&
{\hbar^2 \over 2m} \Bigl\{-\bigl({\cal D}_{F,\mu}\bigr)^2 z_F +z_F {1\over 2} (\partial_\mu \v d)^2
+\sqrt{2}( {\cal D}_{A,\mu}z_A ) \bigl( \v e_{-} \cdot \partial_\mu \v d \bigr) + {1 \over \sqrt 2}
z_A \bigl ( \v e_{-} \cdot \partial^2_\mu \v d \bigr) \Bigr\}\nn & & +{1\over 2} m \omega^2 r^2 z_F
+g_0 \rho z_F + g_2 \rho (z_A^2z_F+z_F^3). \label{geometric eta2 products}\ea
Finally, inner product with $\eta^\dag_{\overline{F}}$ yields

\ba -{i \hbar \over \sqrt 2} z_A  \bigl (\v e_{+} \cdot \partial_t \v d \bigr) &=& {\hbar^2 \over
2m }\Bigl[\sqrt{2} ( {\cal D}_{A,\mu}z_A ) \bigl ( \v e_{+} \cdot
\partial_\mu \v d \bigr)  + {1 \over \sqrt 2}z_A  \bigl( \v e_{+} \cdot
\partial^2_\mu \v d \bigr) + {1\over 2}z_F \bigl(\v e_{+} \cdot
\partial_\mu \v d \bigr)^2 \Bigr]- g_2 \rho  z_A^2z_F.
\label{geometric eta3 products} \ea
\end{widetext}
Here,

\ba {\cal D}_{A,\nu} & \equiv & \partial_\nu + i \partial_\nu \theta+ {1\over 2}\partial_\nu\ln\rho , \nn
{\cal D}_{F,\nu} & \equiv & \partial_\nu + i \partial_\nu \theta + i a_{\nu}+ {1\over
2}\partial_\nu\ln\rho , \label{eq:cov-der} \ea
are the covariant derivatives appropriate for AFM and FM manifold, respectively.

Various hydrodynamic relations existing in the current literature can be derived by going to the FM
limit, $z_A = 0, z_F =1$, and $\v S = \rho \v d$. In this case Eq. (\ref{geometric eta1 products})
is reduced to

\ba i \v e_{+} \cdot D_{F, t} \v d &=& -{\hbar \over 2 m \rho} \v e_+ \cdot \partial_\mu(\rho
\partial_\mu \v d). \label{FM eta1 products} \ea
The velocity vector in the FM manifold is given by $\v v_F = (\hbar/m) (\bm \nabla \theta + \v a)$.
The material derivative, which is different from the covariant derivative given earlier in Eq.
(\ref{eq:cov-der}), then becomes $D_{F, t} \equiv \partial_t + \v v_{F} \cdot \bm \nabla$. The
$g_2$-interaction term vanishes because $\v d \cdot  \v S_{AF} = 0$. By matching the real and the
imaginary parts on both sides of Eq. (\ref{FM eta1 products}) one recovers the Landau-Lifshitz
equation

\ba\rho D_{F,t} \v d &=& {\hbar \over 2 m} \v d \times \partial_\mu(\rho \partial_\mu \v
d) . \label{FM Landau} \ea
Equation (\ref{geometric eta2 products}) in the FM limit has the imaginary part that gives the mass
continuity, $ \partial_t \rho=-\bm \nabla \cdot (\rho \v v_F) $, while its real parts give the
Euler equation~\cite{refael}

\begin{widetext}

\ba
D_{F,t} \v v_F &=& { \hbar \over m} \Bigl[\v v_F \times {\cal B} + {\cal E}
-\bm \nabla \Bigl({ \hbar \over 4 m } \bigl(\partial_\mu \v d \bigr)^2 - { \hbar \over 2m} {\bm \nabla ^2 \sqrt{\rho} \over \sqrt{\rho} } + {1 \over 4} m \omega^2 r^2 + {1 \over 2} g_0 \rho + {1 \over 2}g_2 \rho \Bigr) \Bigr].
\ea
\end{widetext}
Here, ${\cal B} = -\bm \nabla \times \bm a$ and ${\cal E} = \partial_t \bm a - \bm \nabla a_t$ are
effective magnetic and electric fields experienced by the condensate. All the familiar hydrodynamic
relations for the FM manifold are recovered from the first two of the projected equations.

A surprise occurs when we examine Eq. (\ref{geometric eta3
products}), which becomes in the FM limit,

\ba \bigl(\v e_{+}\cdot
\partial_\mu \v d \bigr)^2 &=& 0 \label{FM eta3 products}. \label{FM stuck}
\ea
This is the new and unexpected result found from the third projection. It implies $\partial_\mu \v
d = 0$, and when combined with Eq. (\ref{FM Landau}), also implies $\partial_t \v d = 0$. The
result hold in the presence of the confining trap, as well as the interactions. The only
sustainable dynamics of the $\v d$-vector within the FM manifold is the one of constant, $\v d = \v
d_0$, implying that any inhomogeneity in the initial $\v d$-vector configuration would immediately
throw the condensate wave function $\Psi(\v r , t)$ outside of the FM manifold at $t>0$.

The AFM limit also poses a problem as one can see by examining the limit  $z_A$=$1$, $z_F$=$0$, and
$\v S=0$. This time it is the first projection, Eq. (\ref{geometric eta1 products}), that yields
the mass continuity and the Euler equation:

\begin{widetext}

\ba D_{A,t} \v v_A &=& -{\hbar \over  m} \bm \nabla\Bigl[ {\hbar \over 2 m}(\partial_\mu \v
d)^2-{\hbar \over 2 m}{\bm \nabla^2 \sqrt{\rho}\over \sqrt{\rho}}+{m \over 2} \omega^2 r^2 + g_0
\rho \Bigr]. \ea
\end{widetext}
The velocity field for the AFM condensate is $\v v_A \equiv (\hbar / m) \bm \nabla \theta$.
Definition of the material derivative $D_{A,t}=\partial_t + \v v_A \cdot \bm \nabla$ is similarly
modified. The second projection, Eq. (\ref{geometric eta2 products}), in the AFM limit becomes the
Euler equation

\ba  \rho D_{A,t} \v d &=& -{\hbar \over 2 m} \v d \times \partial_\mu(\rho \partial_\mu \v
d).
\label{L-L s eq AFM} \ea
Finally, the third projection Eq. (\ref{geometric eta3 products}) in the
AFM limit can be re-arranged as

\ba i \rho D_{A,t} \v d
&=& {\hbar \over 2 m} \v d \times \partial_\mu(\rho \partial_\mu \v d) , \label{L-L eq AFM}
\ea
which looks similiar to the LL equation but with an opposite sign on the right side. Combining Eq.
(\ref{L-L s eq AFM}) and Eq. (\ref{L-L eq AFM}), we conclude that each term must vanish separately:

\ba \v d \times
\partial_\mu(\rho\partial_\mu \v d) = 0 , ~ \rho D_{A,t} \v d = 0\label{AFM result}.
\label{AFM restriction} \ea
The first of these results implies

\ba \v d \times \partial_\mu(\rho \partial_\mu \v d) = \partial_\mu
\bigl(\rho \v d \times \partial_\mu \v d \bigr)= 0, \label{eq:2.21}\ea
hence $\rho \v d \times \partial_\mu \v d$ must stay constant throughout the AFM condensate.
If the density $\rho$ was uniform it implies a uniform spiral (or cycloidal) structure for $\v d$. It is the only structures that can sustain dynamics within the AFM manifold. For all other initial configurations the constraint imposed by Eq. (\ref{eq:2.21}) effectively throws the condensate out of the AFM manifold.

Hydrodynamic relations obtained in each specific manifold are arranged in Table \ref{table}.

\begin{table*}
\begin{tabular}{|c|c|c|}\hline
 & AFM & FM \\
\hline
 & $D_{A,t}=\partial_t + \v v_A \cdot \bm \nabla $ &
$D_{F,t}=\partial_t + \v v_F \cdot \bm \nabla $\\
 &$\v v_A = (\hbar / m )\bm \nabla \theta $ & $\v v_F = (\hbar /m )(\bm \nabla \theta + \bm a) $
\\ \hline
$\bm \eta_A^\dag$ & $\partial_t \rho  =  -\bm\nabla \cdot (\rho \v v_A)$ &
$\rho D_{F,t} \v d = {\hbar \over 2 m} \v d \times \partial_\mu(\rho \partial_\mu \v d)$\\
&$D_{A,t} \v v_A = -{\hbar \over  m} \bm \nabla\Bigl[ {\hbar \over 2 m}(\partial_\mu
\v d)^2-{\hbar \over 2 m}{\bm \nabla^2 \sqrt{\rho}\over \sqrt{\rho}}+{1 \over 2}m \omega^2 r^2 + g_o \rho \Bigr]$ & \\\hline
$\bm \eta_F^\dag$ & $ $ & $\partial_t \rho=-\bm
\nabla \cdot (\rho \v v_F)$ \\
& $ \rho D_{A,t} \v d = - {\hbar \over 2m }\v d \times \partial_\mu(\rho \partial_\mu \v d)$& $D_{F.t}\v v_F = {\hbar \over m} \bigl[ \v v_F \times {\cal B } + {\cal E} $\\
 & & $-  \bm \nabla \Bigl( {\hbar \over 4 m } (\partial_\mu \v d)^2-{\hbar \over 2 m }{\bm \nabla^2 \sqrt{\rho}\over \sqrt{\rho}} + {1\over 4}m \omega^2 r^2+{1 \over 2}g_0\rho +{1 \over 2}g_2\rho \Bigr) \bigr]$\\ \hline
$\bm \eta_{\overline{F}}^\dag$ & $\v d \times \partial_\mu\bigr(\rho \partial_\mu \v d \bigr)=0 ,~\rho D_{A,t} \v d =0$ & $\Bigl(\v e_{+} \cdot \partial_\mu \v d \Bigr)^2 = 0$\\\hline
\end{tabular}
\caption{List of hydrodynamic equations obtained in the FM and AFM limits.} \label{table}
\end{table*}

\section{Small fluctuation analysis}
\subsection{Small fluctuation around FM ground state}

We re-examine the previous results~\cite{ho,machida} for the small
fluctuation around the FM ground state in view of the general spin-1
wave function $\Psi_{\mathrm{FM}+\mathrm{AFM}} = \sqrt{\rho} e^{i
\theta} (\bm \eta_F \sin {\delta \over 2} + \bm \eta_A \cos {\delta
\over 2})$. Since the FM ground state is taking place around $\delta
= \pi$, expanding the wave function up to linear order in $\delta$
in the vicinity of $\delta = \pi$ gives
$\Psi_{\mathrm{FM}+\delta\cdot \mathrm{AFM}} = \sqrt{\rho} e^{i
\theta} (\bm \eta_F + \delta \cdot \bm \eta_A) $ where $\delta$
comes from re-defining the mixing angle ${- {\delta \over 2}
\rightarrow \delta}$. We take the fully polarized FM ground state
with the magnetization $\v d$ along the $(0, 0, 1)$ direction. By
noting that the small fluctuation of $\v d = (\cos \alpha \sin
\beta, \sin \alpha \sin \beta, \cos \beta)$ occurs around $\beta =
0$, we can expand the wave function also with respect to small
$\beta$ up to first order:

\ba \Psi_{\mathrm{FM}+\delta\cdot \mathrm{AFM}} \simeq \sqrt{\rho}
e^{i \theta} \left[ \bpm e^{-i (\alpha + \gamma)} \\ {1 \over
\sqrt{2} } \beta e^{-i \gamma} \\ 0 \epm + \delta  \bpm 0 \\ 1 \\
0\epm \right] . \ea
Higher-order terms such as $\delta \times \beta$, $\beta^2$ are
assumed to vanish. The two unit vectors which form a triad together
with $\v d$ are $\v e_x \simeq (\cos (\alpha+ \gamma), \sin
(\alpha+\gamma), 0 )$, $\v e_y \simeq (- \sin (\alpha+ \gamma), \cos
(\alpha+\gamma), 0 )$ for small $\beta$. See Eq. (\ref{eq:ex-ey-d})
for definition of the triad. In the particular analysis at hand the
orientations of $\v e_x, \v e_y$ can be arbitrary without affecting
the physical outcome. In other words, the angle $\alpha + \gamma$
can be chosen arbitrarily. One particular gauge choice $\alpha
+\gamma = 0 $ resulting in $\v e_x \simeq (1, 0, 0)$, $\v e_y \simeq
(0, 1, 0)$ simplifies the above wave function to

\ba \Psi_{\mathrm{FM}+\delta\cdot \mathrm{AFM}} \simeq \sqrt{\rho} e^{i \theta}  \bpm 1  \\ {1 \over \sqrt{2} }d_+ + \delta \\ 0 \epm, \label{eq:FM-flc-wf}\ea
where $d_+  = d_x + id_y= \beta e^{i \alpha}$. The spin average for
the fluctuating wave function is given by $\v S = \Psi^\dag \v F
\Psi = \rho(d_x + \sqrt{2} \delta, d_y, 1)$. Inserting Eq.
(\ref{eq:FM-flc-wf}) into the GP equation gives
\ba && i \hbar \partial_t \left({d_+ \over \sqrt{2}} \!+\! \delta \right) \nn
&& = - {\hbar^2 \over 2 m} \bm \nabla^2 \left({d_+ \over \sqrt{2}}
\!+\! \delta \right) + \rho (g_0 \!+\! g_2) \left({d_+ \over
\sqrt{2}}+ \delta \right).  \nn \ea
We neglect the density fluctuation stemming from the first component
of wave function Eq. (\ref{eq:FM-flc-wf}) since it is massive. We
search for a solution of the form
\ba \left({d_+ \over \sqrt{2}} + \delta \right) = e^{-i \mu t} z_0
(e^{i \v k \cdot \v r - i \omega t} ), \ea
with the overall phase factor $e^{-i \mu t}$. Equating the chemical
potential $\mu$ with $\mu = \rho (g_0 + g_2)$ cancels out the $(g_0
+ g_2)$ term in the dispersion relation and leads to the well-known
quadratic spin-wave dispersion $\hbar \omega_{\v k} = \hbar^2 \v k^2
/2m$~\cite{ho,machida}.
\\

\subsection{Small fluctuation around AFM ground state}

Since the AFM ground state is taking place around $\delta =0$,
expanding the wave function up to the linear order in $\delta$ gives
$\Psi_{\mathrm{AFM}+ \delta\cdot \mathrm{FM}} = \sqrt{\rho} e^{i
\theta} (\bm \eta_A + \delta \cdot \bm \eta_F )$ after the
replacement ${\delta \over 2}\rightarrow \delta$. To describe the
small fluctuation about the AFM ground state,  we also expand each
wave function $\bm \eta_A$, $\bm \eta_F$ for small $\beta$ up to
linear order:
\ba \Psi_{\mathrm{AFM}+\delta\cdot\mathrm{FM}}  \simeq \sqrt{\rho}
e^{i \theta} \left[ \bpm -{1 \over \sqrt{2}}d_-   \\ 1 \\ {1 \over
\sqrt{2}}d_+ \epm  + \delta \bpm e^{-i (\alpha + \gamma)} \\ 0 \\
0 \epm \right]. \nn \label{eq:AFM-fluctuated-wf}\ea

Suppose for a moment that one tried to capture the small fluctuation
effects without leaving the AFM manifold, say by turning off
$\delta$ from the above. Inserting such a wave function into the GP
equation leads to a pair of equations,

\ba i \hbar \partial_t \left({d_- \over \sqrt{2}}\right) &=&
{\hbar^2 \over 2m} \bm \nabla^2 \left({d_- \over \sqrt{2}}\right) ,
\nn
i \hbar \partial_t \left( {d_- \over \sqrt{2}} \right) &=& -{\hbar^2
\over 2m} \bm \nabla^2 \left( {d_- \over \sqrt{2}} \right),
\label{eq:AFM-flc-without-delta}\ea
which are in obvious contradiction to each other, or one must set
each of the terms in the equation to zero, freezing the dynamics
altogether. By allowing the FM component ($\delta \neq 0$), though, the
coupled equations are modified to

\begin{widetext}
\ba  i \hbar \partial_t \left({d_- \over \sqrt{2}}\right) &=&
{\hbar^2 \over 2m} \bm \nabla^2 \left({d_- \over \sqrt{2}}\right) -
g_2 \rho \left(\delta e^{-i (\alpha + \gamma)} \right), \nn
i \hbar
\partial_t \left( -{d_- \over \sqrt{2}} + \delta e^{-i (\alpha +
\gamma)} \right) &=& -{\hbar^2 \over 2m} \bm \nabla^2 \left(-{d_-
\over \sqrt{2}} + \delta e^{-i (\alpha + \gamma)} \right) + g_2 \rho
\left(\delta e^{-i (\alpha + \gamma)} \right) .
\label{eq:AFM-GP-eq}\ea
\end{widetext}
Solutions can be found in the form
\ba - {d_- \over \sqrt{2}} + \delta e^{-i (\alpha + \gamma)} &=& u e^{i \v k \cdot \v r - i \omega t},\nn
 {d_- \over \sqrt{2}} &=& v e^{i \v k \cdot \v r - i \omega t}, \ea
with the coefficients of $u$ and $v$. Solving the matrix problem
\ba \bpm \hbar \omega - {\hbar^2 \v k^2 \over 2m} - g_2 \rho & -g_2
\rho \\ g_2 \rho & \hbar \omega + {\hbar^2 \v k^2 \over 2m} + g_2
\rho \epm \bpm u \\ v \epm = 0 \ea
successfully re-produces the low-energy dispersion $\hbar \omega_{\v
k} = \sqrt{{\hbar ^2 \v k^2\over 2 m} \left({\hbar^2 \v k^2\over 2m
} + 2 g_2 \rho \right)}$ first found in Refs. \onlinecite{ho,
machida}. In both examples of FM or AFM ground states, fluctuation
into the ``other sector" as expressed by non-zero mixing angle
$\delta$ is an inevitable ingredient in the proper dynamical
description.

\section{Numerical Solution of Gross-Pitaevskii Equation}

To simulate the $F=1$ condensate dynamics in one- and two-dimensional systems we must solve the GP equation,

\ba i \hbar \frac{\partial}{\partial t} \Psi &=& \Bigg( - {\hbar^2 \over 2m}\bm \nabla^2 + {1\over
2} m \omega^2  r^2   + g \mu_B B(\v r)F_z   \nn
&& ~ + g_0 \Psi^\dag \Psi + g_2 (\Psi^\dag \v F \Psi) \cdot \v F  \Bigg) \Psi , \ea
where we also included the linear Zeeman term involving the Land$\acute{\text{e}}$ hyperfine
$g$-factor, the Bohr magneton $\mu_B$, and an external magnetic field $B(\v r)$ applied in the
$z$-direction. We employ dimensionless units in which the energy, length, and time scales are
measured by $\hbar \omega$, $\sqrt{\hbar/ m \omega}$, and $1/\omega$, respectively. Here $\omega$
is the frequency of the trapping potential. The dimensionless linear Zeeman term reads $H_z (\v r)
= g \mu_B F_z B (\v r) / \hbar \omega$. The wave function is normalized $\int \Psi^\dag \Psi = 1$
while the total boson number $N$ multiplies the two interaction constants $g_0$ and $g_2$. The GP
equation in dimensionless form becomes

\ba i \frac{\partial}{\partial t} \Psi &=& \Bigg( - {1 \over 2} \bm \nabla^2  + {1\over 2} r^2 +
H_z (\v r)   \nn
&& + {g_0 N \over \hbar \omega}  \Psi^\dag \Psi + {g_2 N \over \hbar \omega} (\Psi^\dag \v F \Psi)
\cdot \v F  \Bigg) \Psi  . \ea
Throughout the simulation we choose $g= -1/2$, and $\omega= 2\pi \times 100$Hz, which makes $H_z
(\v r)  \simeq 6 B(\v r) F_z$/mG. Interaction parameters for the simulation were $g_0 N /\hbar
\omega = 100$ and $g_2 N /\hbar \omega = 10$, respectively.

For the two-dimensional GP simulation, the initial Skyrmion configuration is taken from the $\v
d$-vector

\ba \v d_S = {1\over r^2 \! + \! R^2 } \big(2R y, \, -2R x,
 \, -r^2 \! + \! R^2 \big) , \ea
resulting in the initial-state wave function,

\ba \Psi_A^{(S)} (\v r,t=0) = \frac{\sqrt{\rho(\v r)}}{r^2 \! + \! R^2} \bpm -\sqrt{2} R(ix + y) \\
-r^2 \!+\! R^2
\\ -\sqrt{2} R(ix - y) \epm . \label{eq:2d_AFM_skyrmion}
\ea
The density profile chosen is a gaussian $\rho(\v r) \sim \text{exp}(-r^2 / 2 \sigma^2)$ of width
$\sigma$. The size of the Skyrmion is controlled by $R$. Real-time simulations were performed under
zero and non-uniform ($\v B = B(y) \hat{z}, dB(y)/dy = 1.6 \times 10^{-4}$G/$\mu$m) magnetic fields with $\sigma = 2.2
\mu$m and $R = 3.6 \mu$m, and uniform ($\v B = B_0 \hat{z}, B_0 = 1.6$mG) magnetic fields with
$\sigma = 4.9 \mu$m and $R = 3.6 \mu$m. In all cases, we observe
oscillations of the strong magnetization patterns. Maximum simulation time is set to allow the
observation of a sufficient number of oscillations in the magnetization pattern. Periods of
observed oscillations under zero, uniform, non-uniform magnetic fields were about $6.5 \,
\omega^{-1}$, $6.7 \, \omega^{-1}$, and $6.5 \, \omega^{-1}$, respectively.

For one-dimensional simulation we chose the initial $\v d$-vector configuration
\ba \v d = {1\over x^2 \! + \! R^2 } \big( 2R x, 0, x^2 \! - \! R^2 \big) , \ea
which realizes a rapid rotation of the vector over the length $R$ from the origin. Corresponding
initial-state wave function reads

\ba \Psi_{A}^{(1D)} (x, t=0) = {\sqrt{\rho(x)} \over x^2 + R^2} \bpm - \sqrt{2} R x \\ x^2 - R^2 \\
\sqrt{2} R x \epm, \label{eq:1d_twisted AFM} \ea
where the density profile is a gaussian $\rho(\v r) \sim \text{exp}(-r^2 / 2 \sigma^2)$ of width
$\sigma$. Zero and non-uniform ($\v B = B(x) \hat{z}, dB(x)/dx = 2.0 \times 10^{-4}$G/$\mu$m) magnetic fields were
applied to the initial wave function of $\sigma = 2.0 \mu$m and $R = 2.0 \mu$m. Interaction
parameters for the calculation are the same as in two-dimensional simulation. Again we observe
oscillations of the strong magnetization satellites around the center. Periods of strong
magnetization satellites under zero and non-uniform magnetic fields are about $6.5 \, \omega^{-1}$
for both circumstances.